\documentclass[twocolumn,nofootinbib,preprintnumbers,floatfix,aps,showpacs,prd]{revtex4}

\usepackage[hyperfootnotes=false,bookmarks=false]{hyperref}

\usepackage{graphicx}
\usepackage{amsmath}
\usepackage{multirow}
\usepackage{float}
\usepackage{color}

\newcommand{\be}{\begin{equation}}
\newcommand{\ee}{\end{equation}}
\newcommand{\bea}{\begin{eqnarray}}
\newcommand{\eea}{\end{eqnarray}}

\begin{document}

\title{The twofold emergence of the $a_1$ axial vector meson\\ in high energy hadronic production}

\author{Jean-Louis Basdevant}
\email{jean-louis.basdevant@polytechnique.edu}
\affiliation{Physics Department, Ecole Polytechnique, 91128 Palaiseau, France}
\author{Edmond L. Berger}
\email{berger@anl.gov}
\affiliation{High Energy Physics Division, Argonne National Laboratory, Argonne, Illinois 60439, USA}

\begin{abstract}
The high statistics COMPASS results on diffractive dissociation $\pi N \rightarrow \pi \pi \pi N$ suggest that the isospin
$I=1$ spin-parity $J^{PC}= 1^{++}$ $a_1(1260)$ resonance could be split into two states:  $a_1(1260)$ decaying into an S-wave $\rho\pi$ system, and $a_1^\prime(1420)$ decaying into a P-wave $f_0(980)\pi$ system. We analyse the reaction by incorporating our previous treatment of resonant re-scattering corrections in the Drell-Deck forward production process. Our results show that  the COMPASS results are fully consistent with the existence of a single axial-vector $a_1$ resonance.  The characteristic structure of the production process, which differs in the two orbital angular momentum states, plays a crucial role in this determination. Provided the theoretical analysis of the reaction is done in a consistent manner, this single resonance produces two peaks at different locations in the two channels, with a rapid increase of the phase difference between their amplitudes arising mainly from the structure of the production process itself, and not from a dynamical resonance effect. In addition, 
this analysis clarifies questions related to the mass, width, and decay rates of the $a_1$ resonance. 
\end{abstract}


\maketitle

\section{Introduction\label{sec:int}}

Results of unprecedented precision are emerging from the COMPASS experiment at the CERN SPS where beams of 190 GeV pions interact with proton or nuclear targets, providing new insight into the properties of light mesons~\cite{comp}.  This investigation is of significant interest since it is bound to clarify low-energy hadron spectroscopy, where the situation of some states is somewhat confused compared to expectations in 
a naive view of the standard model. 

In this paper, we concentrate on a question of importance for many reasons: the isospin $1$ axial-vector resonance $a_1$ (reported in the Particle Data Compilation as $a(1)(1260)$~\cite{PDG}.)  Evidence emerges in the COMPASS data for a new narrow $J^{PC} = 1^{++}$ axial-vector state with isospin 1, strongly coupled to the $\pi f_0(980)$ system.  This observation of a  peak in the axial-vector two body  $\pi f_0(980)$ P-wave intensity at a mass of 1.42 GeV, combined with a phase motion close to $180^\circ$ with respect to other waves, appears at face-value to mean that a second axial-vector resonance is present, close in mass to the known broad $a_1(1260)$ that couples mainly to the $\pi \rho$ meson channel \cite{comp}.  While these three features, i.e. two peaks at different masses and a rapid phase variation, are clearly  present,  
there are reasons to be surprised, among which we mention the following:
\begin{enumerate}
\item The $a_1(1260)$ is a central member of the axial-vector nonet, which, together with the $J^{P}= 0^{-},\, 1^{-},\,\textrm{and}\;  0^{+}$ (the latter being somewhat ill-determined yet) form the ground-state of the light quark-antiquark spectrum.
A newcomer in the family would be difficult to accommodate. 
\item It is peculiar to have two $J^{PC} = 1^{++}$ three-pion states, with identical quantum numbers, close in mass (within a full width of each other), with orthogonal decay modes, without the presence of some new quantum number.  The $K^0_S-K^0_L$ system led to decisive discoveries in fundamental physics; neutrino mixing is a spectacular current example.  However, in the $a_1$ case, we see no candidate for a distinguishing quantum number.   

\item In the succesful quark-antiquark potential model approaches to the hadron spectrum, there is no sign that two bound states with identical quantum numbers and comparable masses could be constructed from the $u$ and $d$ quarks (unless the potential is pathological).  One would have to resort perhaps to a four-quark or molecule-type object.

\item Regarding the COMPASS data, we may mention that a similar situation may exist in the case of the production of the $\pi_2(1670)$ and a $\pi_2^\prime$ as s- and d-waves of $\pi f_0(1270)$.  We defer this consideration and report here only on the $a_1$. 
\end{enumerate}

Our basic approach to high energy forward production of three pion states in pion-nucleon interactions is the Drell-Deck model \cite{ddeck}.  This model has been studied extensively in the production of the $J^P=1^+$ $\rho\pi$  system \cite{ELB,ELBJD}, and here we extend the analysis to the $J^P=1^+$ $f_0(980)\pi$  system.  An important difference is that whereas the $\rho\pi$ system is in an orbital S-wave state, the $f_0(980)\pi$ is in an orbital P-wave state, not studied to date.  In addition, since we want to study resonances in these quasi two-body systems, we must modify the Deck mechanism with the proper corrections due to the re-scattering  of these states.  This is an inescapable physical consistency condition of the entire analysis.\\

The unitary coupled channel approach that we developed in the late 1970's in Refs. \cite{BB1}, \cite{BB2} and \cite{BB3}, should be an ideal way to show whether one resonance can produce mass peaks at different locations in the two decay channels, along with a relative phase variation between the two channels, or whether the COMPASS data do require two nearby resonances with the same axial-vector quantum numbers in the three-pion system.  In this paper, we demonstrate that the main features of the COMPASS data are compatible with a \emph{single} resonance.  We concentrate on developing the theoretical method of analysis and its main consequences,  leaving a detailed fit to the data to future discussion with the COMPASS collaboration.  

As we show, the double peak structure, by itself, requires a revision in the determination of the nominal second-sheet pole parameters of the $a_1$, i.e. its mass and width.  Owing to its  large width, the $a_1$ peak, as observed in various final states, appears distorted by several effects.   In diffraction dissociation, as we have shown in \cite{BB1}, the structure of the Deck amplitude alters the resonance peak considerably.  In $\tau$-decay \cite{BB4}, a distortion arises from phase space factors. Therefore, the apparent ``mass'', as identified by the peak position, can vary considerably according to the production mechanism.  As a byproduct of our present study, we are able to determine a new estimate of the branching ratio of the $a_1$ into $\pi f_0(980)$. \\

In Sec. \ref{sec:Deck}, we recall the basic facts about the Deck mechanism.  We exhibit its P-wave structure which has not received much attention before now.  We elicit a simple property that turns out to be the cause of the rapid rise up of the $f_0\pi$ phase relative to $\rho\pi$.\\

Since we are interested in the $a_1$ which corresponds to resonant  behavior of the $\rho\pi$ and $f_0\pi$ amplitudes, we must take into account the final state interactions between these particles and/or the re-scattering corrections to the bare Deck amplitudes.  It would be inconsistent physically not to perform such an analysis that incorporates unitarity requirements.  We published such work in the late 1970's on the $a_1$ and on the $K1(1270)$-$K1(1400)$ system, but our numerical results then were compared with data having much smaller statistics.  In Sec. \ref{sec:unitarization}, we describe the unitarization procedure, and we introduce the relevant physical parameters.  We explain how we deal with a very restricted set of parameters when we introduce a single $a_1$ resonance.

In Sec. \ref{sec:results} we present our results together with the values of physical parameters involved, and we summarize our conclusions in Sec. \ref{sec:summary}. For the sake of clarity we collect the simple but necessary formulae in the Appendix.  

\section{Two-channel Deck amplitudes}
\label{sec:Deck}
\begin{figure}[h]
  \begin{center}
  \includegraphics[width=0.5\textwidth]{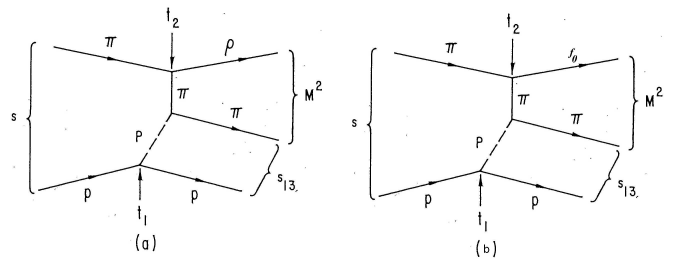}
  \caption{Deck production processes for (a) $\rho\pi$ and (b) $f_0\pi$. }
\label{fig:deckf}
\end{center}
\end{figure}

We follow closely Refs.~\cite{ELB}, \cite{ELBJD} and \cite{BB1}.
We consider the Deck amplitudes $T_D(\pi \textrm{N} \to \pi \pi \pi \textrm{N})$ for production of the $\pi \pi \pi$ system at 
small momentum transfer and high incident energy (known as ``diffractive production").   For the quasi-two body systems $\pi \rho$ and 
$\pi f_0$, we denote  
$$T^{\rho}_D = T_D(\pi \textrm{N} \to \pi \rho \textrm{N})\,\quad\textrm{and}$$
$$T^f_D = T_D(\pi \textrm{N} \to \pi \textrm{f}_0\, \textrm{N})\quad.$$
The reactions are represented in Figs.\ref{fig:deckf} (a) and (b). 

The $\pi \rho$ case has been studied at length. Its amplitude is given by Eq.~(2.1) of Ref.~\cite{BB1} as
\begin{equation}
T^{\rho}_D= g_{\rho\pi\pi}K_\rho(t_2)\frac{1}{m^2_\pi -t_2} {i}s_{13}e^{b t_1}\sigma_{\pi p}
\label{DCK}
\end{equation}
where $g_{\rho\pi\pi}$ is the $\rho\pi\pi$ coupling constant ( $g^2/(4\pi) =2.4$), $K_\rho$ is the magnitude of the incident pion momentum in the 
$\rho$ rest frame~\cite{ELBJD}, $b$ is the slope of the $\pi N$ elastic 
diffraction peak, and $\sigma_{\pi p}$ is the $\pi p$ total cross-section. The invariants
$s_{13}$, $t_1$ and $t_2$ are labeled in Fig.~\ref{fig:deckf}~(a).\\

Similarly, the $\pi f_0$ production amplitude is 
\begin{equation}
T^{f}_D= g_{f_0\pi\pi}\frac{1}{m^2_\pi -t_2} {i}s_{13}e^{b t_1}\sigma_{\pi p}\quad.
\label{FCK}
\end{equation}

This expression must be projected onto the orbital angular momentum $l = 1$ P-wave of the $\pi f_0$ system for the 
purposes of our present investigation.  Note here that the $f_0\pi\pi$ coupling constant, $g_{f_0\pi\pi}$, has the dimension 
of a  momentum.  Choosing the average value of the 
$f_0 \rightarrow \pi\pi$ width of $60$~MeV, we obtain a numerical value $g_{f_0\pi\pi} \simeq 1.45$~GeV. The other factors 
in Eq.~(\ref{FCK}), relative to  Fig.~\ref{fig:deckf}~(b) have the same meaning as in Eq.~(\ref{DCK}). 

\subsection{Background $\rho\pi$ Deck amplitude}
The Deck background amplitude has been well studied, where,  by background, we mean the amplitude before any unitarization or rescattering correction is performed.  We refer to our previous work in Ref.~\cite{BB1} and extract what is useful in the present analysis. 
We work in the final $\rho\pi$ (or, more generally, resonance-$\pi$) center of mass frame; $M$ is the invariant mass of this system, and we denote 
by $\theta$ the angle between the incident and outgoing pion momenta. 

In the limit of forward production $(t_1\to 0)$ and large $s$, the amplitude in Eq. (\ref{DCK}) takes on a very simple form; see, 
e. g., Ref~\cite{ELB}
\begin{equation}
T^{\rho}_D (M^2, s, t_1=0)\simeq \alpha \frac{i{s}}{M^2 -m^2_\pi}\quad, 
\label{Dt10}
\end{equation}
which is nothing but the ($\rho\pi$) S-wave projection $T^{\rho}_{J^P=1^+}$ of Eq.(\ref{DCK}), where $\alpha$ is a known constant that we discuss below.  This S-wave projection has been used in previous calculations (e.g., Refs~\cite{BB1} and \cite{BB2}).   

In the limit $t_1\to 0$ and large $s$, the Deck amplitudes for higher partial waves vanish identically.  
However, the $J^P=1^+$ $f_0\pi$ system is in an orbital P-wave.  To address $f_0\pi$, we must extend the partial wave extraction calculations to 
finite values of $t_1$ and $s$.  The complete calculation of these amplitudes is presented in the Appendix.  The important feature is that the higher partial wave amplitudes are of order $t_1/M^2$ or $M^2/s$ compared to Eq.~(\ref{Dt10}), and this S-wave amplitude is modified slightly.  An immediate consequence is that $f_0\pi$ P-wave production should have a noticeably smaller rate than the $\rho\pi$ S-wave process, 
as is borne out in the complete calculation in the Appendix, and exhibited by the COMPASS data, where the intensity of the $f_0\pi$ peak at 1.42\,GeV, is lower than that of the  $\rho\pi$ peak at 1.26\,GeV by a factor of the order of a few $10^{-3}$.\\

In the COMPASS experiment Ref~\cite{comp}, the value of the square of the invariant total energy is 
$s= 380$~GeV$^2$ while the momentum-transfer $t_1$ in the smallest bin is $t_1 \in [-0.1,-0.13]$~GeV$^2$
(we follow the experimenters' definition $t_1= t-t_{min}$, except that, for our convenience, we work with negative momentum transfers).  In this analysis, we are interested in values of $M\sim 1 $ to $2$~GeV. We notice that $\vert t_1 \vert/ M^2 \gg M^2/s$ and therefore the only relevant kinematic corrections come from the momentum transfer dependence (confirmed in the quantitative analysis in the Appendix).  We choose to work at the fixed value of $t_1=-0.1$~GeV$^2$, and we checked that within the first t-bin ($t_1 \in [-0.1,-0.13]$~GeV$^2$), our results do not vary appreciably.

\subsection{$\rho\pi$ and $f_0\pi$ background amplitudes to first order}
In the Appendix we derive expansions of the background amplitudes to first order in the momentum transfer.  A convenient dimensionless 
expansion parameter is

\begin{equation}
\Theta_1 = \frac{t_1}{(M^2-m_{\pi}^2)} .
\label{tetard}
\end{equation}

The $J^P=1^+$ S-wave $\rho\pi$ background amplitude  is, to first order in $\Theta_1$,
\bea
T^{Deck}_S&=& -\frac{s}{(M^2-m^2_\pi)}\times \nonumber\\ 
&&\left( 1-\frac{1}{2}{\Theta_1}(\frac{(3M^2+m^2_\pi)}{(M^2-m_\pi^2)}- \frac{E_{\rho}}{E_\pi})(\frac{1}{y}\ln\frac{1+y}{1-y})
\right)
\quad,
\label{ropib}
\eea

where $E_\pi$ and $E_\rho$ are the pion and  $\rho$ energies in the 
 $\rho\pi$ rest frame, and where 
 \begin{equation}
y= p_\pi/E_\pi,
\label{ygrec}
\end{equation}
is the $\rho\pi$ phase space factor, $p_\pi$ being the pion momentum in the $\rho\pi$ rest frame.

The  $J^P=1^+$ P-wave $f_0\pi$ amplitude is, at the same order in $\Theta_1$,
\bea
 T^{Deck}_P&=&+\frac{3}{2}\frac{s}{(M^2-m^2_\pi)} {\Theta_1}\times \nonumber\\
&&\left(\frac{(3M^2+m^2_\pi)}{(M^2-m^2_\pi)}- \frac{E_{f_0}}{E_\pi}\right) (\frac{-2}{y} +\frac{1}{y^2}\ln(\frac{1+y}{1-y}))
\quad, \label{fop}
\eea
where $E_\pi$, $E_{f_0}$ are the pion and $f_0$ energies, $p_{\pi}$ the pion momentum in the $f_0\pi$ rest frame and, as above, $y=p_{\pi}/E_\pi$.\\

\begin{figure}[h]
\begin{center}
{\includegraphics[width=.45\textwidth]{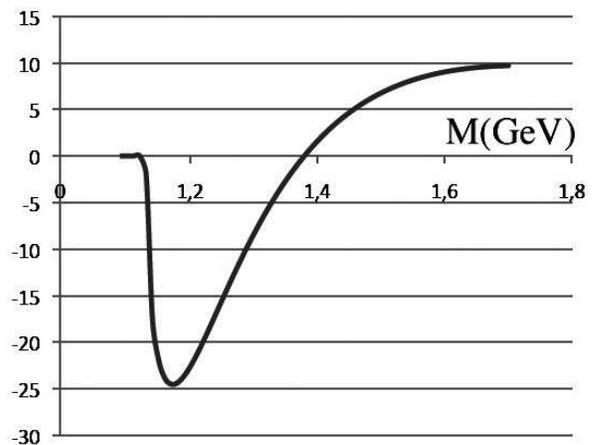}}
\caption{Behavior of the background P-wave $f_0\pi$ Deck amplitude above threshold, showing the zero of this amplitude near $1.38$~GeV.}
\label{fig:foz}
\end{center}
\end{figure}
This latter expression contains a major clue to our investigation.  Of course, we shall unitarize these formulas, and combine them properly, but the simple expression of this 
$f_0\pi$ Deck amplitude explains an important feature of the data. Indeed, the right hand side contains the factor 
${(3M^2+m^2_\pi)}/{(M^2-m^2_\pi)}- {E_{f_0}}/{E_\pi}$,
plotted in Fig.~(\ref{fig:foz}).  This factor is negative at low values of $M$ (since $m_{f_0}/m_\pi > 3)$, but it vanishes at some point near $M\simeq 1.38$~GeV and becomes positive afterward.  Furthermore, if we give this term some small imaginary part, its phase will suddenly switch from $-180^\circ$ to zero.  This sudden and rapid phase variation is not a dynamical effect in the sense of a resonant phase, but it originates in the structure of the dynamical process by which the $f_0\pi$ state is produced.  A similar term is present also in the S-wave, Eq. (\ref{ropib}), but there it is completely hidden by the dominant term.\\

Another interesting qualitative feature of Eq.~(\ref{fop}) is that it grows in the region of interest ($M\sim 1.2 $ to $1.4$~GeV) and therefore tends to push a resonance peak upward in $M$, and, because of the zero, to produce a (non-resonant) peak or maximum between threshold and 
$M= 1.38$~GeV.

\subsection{Parameters and normalization}
Keeping in mind the parameters introduced in Eqs.~(\ref{DCK}) and (\ref{FCK}), our two $J^{PC}=1^{++}$  amplitudes are 
\begin{equation}
\left( \begin{array}{l}
T_{Deck} (\rho\pi)\\
T_{Deck} (f_0\pi)
\end{array}\right) = \frac{2i\sqrt{2}s N}{(M^2-m_\pi^2)}\left( \begin{array}{l}
g_{\rho\pi\pi} K_\rho\sigma_{\pi p} \tilde T_{\rho\pi}\\
g_{f_0\pi\pi}\;\sigma_{\pi p}\; \tilde T_{f_0\pi}\end{array}\right) ,
\label{tdkrf}
\end{equation}
where, $\tilde T_{\rho\pi}$ and $\tilde T_{f_0\pi}$ can be read off from Eqs.~(\ref{ropib}) and (\ref{fop}).   The structure remains the same after 
we unitarize.  The normalization factor $N$ is irrelevant and is taken equal to $1$ here.  In a more complete analysis $N$ will ensure precise quantitative fits to the data.\\

After integration over phase space, the normalized differential cross sections, expressed in GeV units, are 
\begin{equation}
\frac{d\sigma}{dt_1dM}\vert_{t_1}= \frac{1}{2}\frac{1}{0.3893}\frac{q_{a^*\pi}}{(4\pi)^3 s^2}\vert T_{Deck}(a^*\pi)\vert^2\quad,
\label{crosect}
\end{equation}
where $a^*$ stands for $\rho$ or $f_0$, and $q_{a^*\pi}$ is the magnitude of the $a^*$ in the $a^*\pi$ rest frame.

\section{Unitarization}
\label{sec:unitarization}
For all theoretical and technical details about multi-channel final state unitarization, we refer to the literature, in particular to Ref.~\cite{BAB} where the general analysis is done thoroughly and to Ref.~\cite{BB1} where a specific application is made.
Let us simply recall that if $\mathcal {S}$ is the (two-channel) strong interaction $S$ matrix, then the unitarized  Deck amplitude $\mathcal {T_D}$, which we can write as a two dimensional vector as in Eq.(\ref{tdkrf}), has a right hand unitarity cut along which it satisfies the relation 
\begin{equation}
\mathcal{T_D}^+ = \mathcal S \mathcal{T_D}^-
\label{unita}
\end{equation}
$\mathcal{T_D}^+$ and $\mathcal{T_D}^-$ being the values of the unitarized Deck amplitude above and below the cut.

\subsection{The resonant $\rho\pi$ and $f_0\pi$ systems}
In order to unitarize the Deck amplitude by taking into account resonant inelastic (or coupled channel) final state interactions between the $\rho\pi$ and $f_0\pi$ states, we introduce Chew-Mandelstam functions~\cite{CM} $C^\rho$ and $C^f$.  By definition these are analytic functions of the invariant energy squared $s$ of two particles, with a right hand branch cut where the imaginary part is equal to the phase space factor $2p/\sqrt{s}$, $p$ being the c.m. momentum:
\bea
&&C_{m,\mu}(s) \equiv C(s;m,\mu) = \nonumber\\
&& -\frac{2}{\pi}{\large\lbrace}-\frac{1}{s}[(m+\mu)^2-s]^{1/2}[(m-\mu)^2-s]^{1/2}
\nonumber\\
&& \times\ln\frac{[(m+\mu)^2-s]^{1/2}+[(m-\mu)^2-s]^{1/2}}{2(m\mu)^{1/2}}\nonumber\\
&&
 +\frac{m^2-\mu^2}{2s}\ln\frac{m}{\mu}- \frac{m^2+mu^2}{2(m^2-mu^2)}\ln\frac{m}{\mu}
-\frac{1}{2}{\large\rbrace}.\
\label{fcm}
\eea
The function $C^\rho(M^2)$ (where $M^2$ is the square of the c.m. energy) is written at length in Eq.~(3.8) of Ref.~\cite{BB3};  $C^f$ is given by a slightly different formula since $\pi f_0$ is in a P-wave. \\

The P-wave nature has a few simple consequences.  An elastic $\pi f_0 \to\pi f_0$ amplitude behaves, at threshold, as
\begin{equation}
p_{\pi f_0}^2(M^2)= \frac{[M^2-(m_f+m_\pi)^2][M^2-(m_f-m_\pi)^2]}{4M^2}\quad,
\label{pdeux}
\end{equation}
and an inelastic amplitude such as $\pi \rho \to\pi f_0$ behaves as $p_{\pi f_0}(M^2)$.  This last property can be seen in the Deck production amplitude Eq.~(\ref{fop}) which is proportional to $y=2p/M$ as $y\to 0$.

In practice, since Eq.~(\ref{fcm}) vanishes at $M^2=0$, and therefore $p_{\pi f_0}^2(M^2)\times C(M^2;m_f,m_\pi)$ is regular and analytic in the cut plane from $(m_f+m_\pi)^2$ to infinity with a discontinuity equal to $2p^3/M$, it is more convenient for us to work with the two Chew-Mandelstam functions of different dimensions:
\begin{equation}
C_1(M^2)\equiv C^\rho(M^2) = C(M^2;m_\rho,m_\pi)\quad,
\label{C1}
\end{equation}
and
\begin{equation}
C_2(M^2)\equiv C^f(M^2) = p_{\pi f_0}^2(M^2)C(M^2;m_f,m_\pi)\quad.
\label{C2}
\end{equation}
These have different dimensions, but they will appear more symmetrically in the formulas, making these easier to read. They are also associated with constants of different dimensions, and we take care, inside the complete formulas, to check the presence of the appropriate powers of the momenta. This is an alternative to the more general treatment described at length in the literature, for instance in Eqs. (3.1) to (3.4) of Ref.~\cite{BAB}.

\subsection{Unitarization procedure}
We follow the extensive coupled channel unitarization procedure presented in Sec. III of Ref.~\cite{BB1}.
We assume that the $\rho$ and the $f_0$ are approximately stable.  Therefore $C^\rho$ and $C^f$ are defined analytically as above.  
Furthermore, we assume that the $f_0$ is an {{elastic}} $\pi\pi$ resonance.  We know that this is not true.   There is a noticeable $K\bar{K}$ inelasticity,  but it is not predominant.  If it turns out to be necessary to incorporate inelasticity, the simplest procedure consists in introducing the $K\bar{K}$ partial decay mode through a third channel $K\bar{K}$, with the corresponding  Chew-Mandelstam function $C^K$.\footnote{We actually built an unstable Chew-Mandelstam function for the $f_0$ following what we did in Ref.~\cite{BB2} in the case of the $K1(1270)$ and $K1(1400)$ strange axial vector mesons. This procedure reproduces the decay width of the $f_0$ as well as the observed $\pi-\pi$ inelasticity~\cite{JLBMP} at the opening of the $K\bar K$ channel, in analogy with the empirical Flatt\'e formula \cite{FLAT}. However, we found very small deviations in the results compared to the stable case because the coupling of the $a_1$ resonance to the $f_0\pi$ channel is small.  We might restore inelasticity in a subsequent 
analysis.}\\

Our basic assumption, in the present analysis, is that there is \emph{a single $a_1$ resonance whose (unique) second-sheet pole parameters we shall determine}.
Since we are dealing with a two-channel case, we parametrize the coupled $\rho\pi$ and $f_0\pi$ final state interactions (or rescattering) via this resonance.\\

In order to do this, we introduce a $K$ matrix, as in Eq.(3.14) of \cite{BB1}  
\begin{equation}
 K(M^2)=\left( \begin{array}{cc}
\frac{g_1^2}{s_1-M^2}&\frac{g_1g_2}{s_1-M^2}\\
\frac{g_1g_2}{s_1-M^2}&\frac{g_2^2}{s_1-M^2}\\
\end{array} \right) \quad.
\label{Kmat}
\end{equation}
As mentioned above, since the two channels have the same total angular momentum but different orbital angular momenta, the $a_1\to\rho\pi$ coupling constant $g_1$ has the dimension of an energy, while the $a_1\to f_0\pi$ coupling constant $g_2$ is dimensionless. The squared mass $s_1$ will be related to the $a_1$ resonance mass.\footnote{{One can perfectly well operate in a more standard way, but the momentum factors generate slightly complicated formulas. Here, one can include the $p$ factors due to the $\pi-f_0$ amplitude directly into the constant $g_2$ which would become an expression of the type $p\,\tilde g_2$. This would ensure the correct P-wave behavior of the $\pi-f_0$ channel, without having to change the $C^f$ function.}} All our units are in GeV.  We work with the parameter $\gamma= g_2/g_1$ in discussing the results.
\\

The crucial tool to treat coupled channel final state interactions is the $D$-matrix, which, in this case, is given explicitly in Eq.(3.15) of \cite{BB1} by
\begin{equation}
D(M^2)= \frac{1}{\mathcal{D}_0(M^2)}\left( \begin{array}{cc}
g_1 & -g_2(s_1-M^2-\alpha^2C_2)\\
g_2 & g_1(s_1 -M^2- \alpha^2 C_1 \end{array}\right)
\label{Dmat}
\end{equation}
where $\alpha^2= g_1^2+g_2^2$,  and where the energy denominator function $\mathcal D_0(M^2)$ is 
\begin{equation}
\mathcal D_0(M^2)= (s_1-M^2- g_1^2 C_1(M^2)-g_2^2 C_2(M^2))\quad.
\label{Dfunc}
\end{equation}

\begin{figure}[h]
\begin{center}
{\includegraphics[width=.45\textwidth]{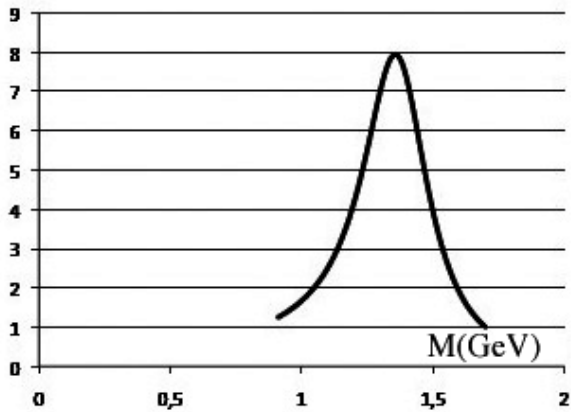}}
\caption{Behavior of  $1/{\vert \mathcal D_0(M^2)\vert}^2$ as a function of the energy $M$, in the case
of an input second sheet pole corresponding to $m_{a_1}= 1.4$~GeV and $\Gamma_{a_1}=0.3$~GeV as used in this analysis.}
\label{D0}
\end{center}
\end{figure}

The function $\mathcal D_0(M^2)$  is of fundamental importance. It contains all the information (that we put in) on the coupled-channel $\rho\pi$-$f_0\pi$ strong interaction. Equation (14) of Ref.~\cite{BB2} shows how the matrix $D$ gives direct access to the strong interaction $t-$matrix (whose first diagonal element is $e^{i\delta}\sin \delta$ in the elastic region). $\mathcal D_0(M^2)$ is an analytic function which possesses the $\rho\pi$ and $f_0\pi$ branch cuts from $[m_\rho+m_\pi]^2$ to infinity and from $[m_{f_0}+m_\pi]^2$ to infinity. Its second sheet  pole determines the nominal position and width of the $a_1$ resonance.  The function $\mathcal D_0(M^2)$ is illustrated in Fig.~(\ref{D0}).\\

As in \cite{BB1}, we \emph{define the unitarized Deck amplitude} as 
\bea
T_D^u(M^2)&=& T_D(M^2) - \frac{1}{\pi}D(M^2)\times \nonumber\\
&&\int_{(m_\rho+m_\pi)^2}^\infty\, ds'\,\frac{ImD(s') T_D(s')}{(s'-M^2)}\quad.
\label{tdu}
\eea
Here $T_D^u(M^2)$ is a two-dimensional vector and $T_D(M^2)$ is the ``background" Deck amplitude discussed above.\\

Equation (\ref{tdu}) is the Deck amplitude with rescattering corrections taken into account. It has the same left-hand singularities as $T_D(M^2)$ (for instance the pole at $M^2=m_\pi^2$), it satisfies the unitarity relation Eq.~(\ref{unita}).  It reduces to $T_D(M^2)$ if the strong interaction amplitudes vanish. In the Appendix of Ref.~\cite{BB1} we justify this interpretation qualitatively within a simple effective Lagrangian model.

\subsection{Direct production amplitude}

\begin{figure}[h]
\begin{center}
{\includegraphics[width=.4\textwidth]{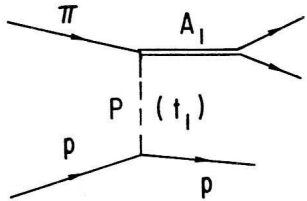}}
\caption{Diagram representing direct production of $a_1$ resonance through coupling to the diffractive exchange mechanism (pomeron.}
\label{tdp}
\end{center}
\end{figure}

There is one more ingredient.  In the same effective Lagrangian model mentioned above, we show that in most cases, such as the present problem, one must also incorporate a \emph{direct production} contribution of the form
\begin{equation}
T(M^2)= T_D^u(M^2) +D(M^2)P(M^2) \quad,
\label{tdd}
\end{equation}
where $P(M^2)$ is a constant or first degree polynomial. 
The corresponding process is illustrated in Fig.~(\ref{tdp}).

There are various arguments in favor of this term, such as high energy behavior.    
The most intuitive is based on an examination of a one channel case where the explicit result, the Omn\`es-Mushkhelishvili equation~\cite{OMSK}, 
has been known for a long time.  If the resonance is not a dynamical effect of the forces, but rather a CDD pole, i.e. a system with appropriate quantum numbers whose origin comes from another system, such as a quark-antiquark state, then a form of the type of Eq.~(\ref{tdu}) will produce a unitarized amplitude behaving as $e^{i\delta}\cos\delta$ with a zero near where we expect the resonance to appear ($\delta\sim 90^\circ$), whereas the direct production term  behaves as $e^{i\delta}\sin\delta$ with a bona fide peak.  A combination of the two terms is necessary in general.  When we display the various stages of $T_D^\rho$ before and after unitarization, we see how this zero appears, why direct production is necessary by itself, and also why both contributions are necessary in order to build the observed resonance peak.

In the present analysis, the direct production term must also be diffractive and obey the unitarity constraints. We choose an amplitude of the form
\begin{equation}
T_{dir}(s,M^2) = \frac{is \sigma_{\pi p}G}{\mathcal D_0(m^2)}\left(\begin{array}{l}
f_1 \\ f_2\end{array} \right) ,
\label{dirprod}
\end{equation}
where $G$ is an arbitrary constant.
Ideally, one would expect the two constants which appear here $f_1$ and $f_2$ 
to be proportional respectively to $g_1$ and $g_2$, which are the couplings of the $a_1$ to the $\rho\pi$ and $f_0\pi$ channels. This proportionality turns out not to be possible, although the values we find have appropriate orders of magnitude (i.e. $\vert f_2\vert \ll \vert f_1\vert$).  However, as we have pointed out previously~\cite{BB2}, there is arbitrariness in the definition of the direct production amplitude, and it is not of primary importance for our main conclusions.

\section{Analysis and results}
\label{sec:results}
In tis section we present the main results of our investigation.  We have \emph{not} attempted a detailed comparison with data, based, for 
example on a minimization procedure.  Nevertheless, there are some salient points that can be made.\\   

Our first ambition has been to select appropriate values of the $a_1$ parameters that provide a good global description. This goal was not as trivial 
as we thought.  In our previous paper on the $a_1$ \cite{BB1} we had been able to explore a wide range of values of the $a_1$ mass and width. It turns out that the COMPASS results fix these parameters in a more stringent way than when we dealt only with the S-wave $\rho\pi$ system, be it or not associated with other S-wave channels (such as $K^* \bar K$).  Here, the acceptable mass and width of the $a_1$, \emph{as defined by the position of the second sheet pole}, which are the actual parameters of our calculation (i. e., after fixing them, we determine the values of $s_1$ and $g_1$ in Eq.(\ref{Kmat})), turn out to be quite restricted.  Despite the fact that it is more oriented towards qualitative aspects of the COMPASS data than to quantitative fits, our analysis indicates that:
\begin{center}
$M(a_1) \simeq 1.40\pm 0.02~{\rm {GeV}}$ \\

$\Gamma(a_1) \simeq 0.30\pm 0.05~\rm {GeV}$. 
\end{center}
These values of mass and width correspond to values of the  parameters $s_1 \sim  2.002$~GeV$^2$ and $g_1 \sim   0.732$~GeV.  We have varied the parameters in order to understand the global trends, but in this first calculation we prefer to keep them fixed: $M(a_1)= 1.4$~GeV, $\Gamma(a_1)= 0.3$~GeV (within the uncertainty ranges noted above). \\

The  interesting parameter to vary is the ratio $\gamma=g_2/g_1$ in order to find the range of values that produce two peaks with appropriate characteristics:  the $f_0\pi$ peak occurs at higher mass than the $\rho\pi$ peak, and the ratio of maximal intensities of these peaks, i.e. 
$\rho\pi$/$f_0\pi$, falls between 1,000 and 500, as indicated by the available data. 
These requirements lead to \emph{negative} values of $\gamma$ in the range $[-0.1, -0.055]$. In other words, the $a_1$ couplings to $\rho\pi$ and to $f_0\pi$ have opposite signs!  The fact that $\gamma$ is negative is quite easy to understand. If we look at Fig.~(\ref{fig:foz}), we see that the relative sign of the P-wave and S-wave amplitudes determines whether the phase switches from $-180^\circ$ to $0$ , i.e. upward, or from $0$ to  $-180^\circ$, i.e. downward.  For the same reason, the P-wave peak will tend to move to higher energies than the S-wave peak.  In other words, the data impose that the relative couplings have opposite signs.  We choose as our central value
$$\gamma=g_2/g_1= -0.08\;,$$ 
and we show how the energy variation of the phase changes with this parameter.\\

The next step is to determine the amount of direct production necessary to fix the two peaks, one in $\rho\pi$, the other in $f_0\pi$, at their desired positions, i.e. $M=1.26$~GeV for  $\rho\pi$ and $M=1.42$~GeV for $f_0\pi$. 
Values such as $$G\sigma_{\pi p} f_1 = 120\quad \textrm{and} \quad G\sigma_{\pi p} f_2= 5.5$$ 
in Eq.(~\ref{dirprod}) ensure good positions for the two peaks, and this situation is stable when one varies the parameter $\gamma$.\\
Two facts seem important.  First, the parameters which are essential in fixing accurately the positions of the two peaks are the direct production amplitudes above.  It is not possible to change appreciably the position of either of the peaks by varying the parameter $\gamma$.  Some variation is 
possible if one changes the mass and width of the $a_1$, but this approach leads to a deviation from the desired values of $1.26$ and $1.42$ respectively. 
The second point is that the ratio of direct production to the background Deck amplitude is consistent with the value we obtained some time ago \cite{BB1} in our analysis of data at much lower energies, with smaller statistics (see Ref.~\cite{AA}).  In our opinion, this is a positive outcome. In fact, in the present analysis, the inclusion of the $f_0\pi$ channel is a small perturbation on the dominant S-wave $\rho\pi$ amplitude, and it is satisfactory to see that the results confirm this.\\

\begin{figure}[h]
\begin{center}
{\includegraphics[width=.5\textwidth]{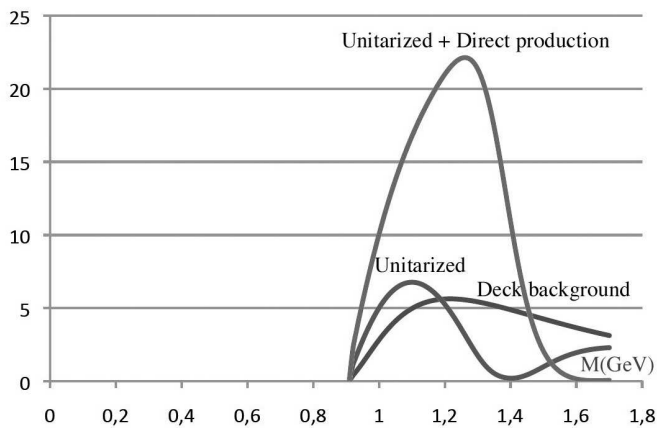}}
\caption{Evolution of the $\rho\pi$ differential cross section as a function of the energy $M$, in three cases: Background Deck, Unitarized Deck as in Eq.(\ref{tdu}), and final version including direct production. }
\label{rhovs}
\end{center}
\end{figure}

It is interesting to see how the two peaks build up, now that we know in what range we can choose the parameters. 
We plot in Fig.~(\ref{rhovs}) the shapes of the rho intensity as a function of the energy $M$, for the various stages of the calculation.
We notice that the pure Deck background does not produce a resonant shape. The unitarized amplitude shows effectively the $\sim \cos\delta$ zero we mentioned above in the one-channel case (to which this problem is actually very close). Finally, the direct production produces the observed peak, at the observed position.\\

\begin{figure}[h]
\begin{center}
{\includegraphics[width=.5\textwidth]{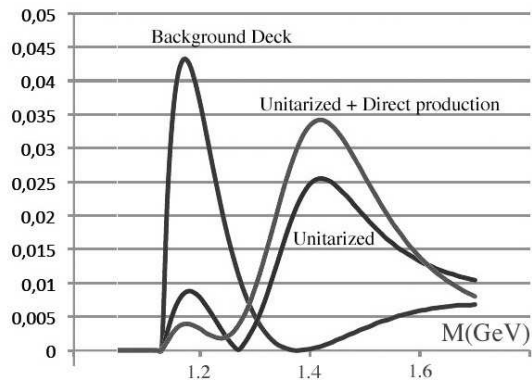}}
\caption{Evolution of the $f_0\pi$ differential cross section as a function of the energy $M$, in three cases: Background Deck, Unitarized Deck as in Eq.(\ref{tdu}), and final version including direct production. }
\label{f0vs}
\end{center}
\end{figure}

A similar set of curves for the  $f_0\pi$ intensity is shown in Fig.~(\ref{f0vs}). Notice that the form of the Deck background by itself appears to simulate a narrow resonance peak at threshold (of course without any accompanying phase).

\begin{figure}[h]
\begin{center}
{\includegraphics[width=.45\textwidth]{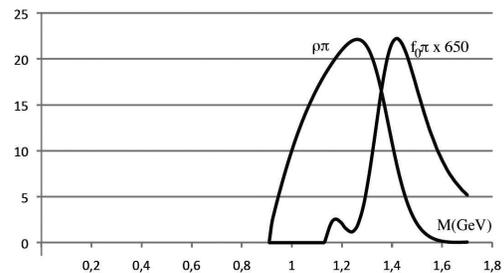}}
\caption{$\rho\pi$ and $f_0\pi$ differential cross sections as a function of mass.  The second is multiplied by a factor of 650 to make the figure easier to understand.}
\label{rof0}
\end{center}
\end{figure}

Because the $f_0\pi$ peak has a small intensity, in order to see properly the separation of the S-wave and P-wave peaks, we prefer to 
multiply the $f_0\pi$ intensity by an appropriate coefficient of 650 such that it appears on an equal footing with the $\rho\pi$ peak. This result 
is shown in Fig.~ (\ref{rof0}).  

The separation of the positions of the two peaks is evident.  The width of $a_1(1260)$ is about twice the width of $a_1(1420)$ in the calculation, in agreement with the COMPASS observation.  The $a_1(1420)$ peak is also more symmetrical, with width about $0.14$~GeV.  One interesting detail is that the lower end of the $f_0\pi$ intensity exhibits a (tiny) peak at around $1.2$~GeV owing to the zero emphasized in 
Fig.~(\ref{fig:foz}), which in turn is part of the dynamical structure of the P-wave Deck amplitude that is responsible for the sharp rise of the phase difference between the $f_0\pi$ and $\rho\pi$ amplitudes.  Perhaps this small peak near $f_0 \pi$ threshold is accessible in the experiment. \\

\begin{figure}[h]
\begin{center}
{\includegraphics[width=.45\textwidth]{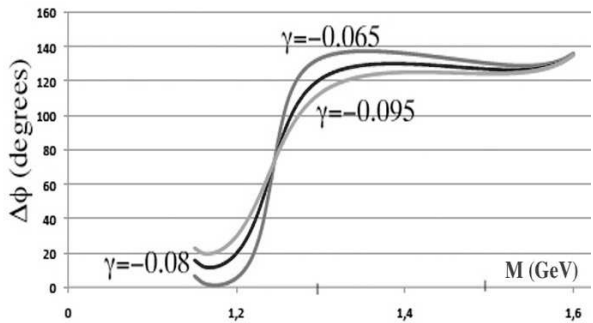}}
\caption{Three phase-differences between $f_0\pi$ and $\rho\pi$, in the 1.2 to 1.6 GeV region. The medium curve corresponds to our ``central" solution $\gamma=-0.08$ the two others (marked) to other values in the range of interest. In all three cases, the other results of the calculation remain practically unchanged: peak positions and intensities.}
\label{phas}
\end{center}
\end{figure}

We display in Fig.~(\ref{phas}) a set of phase differences between the $f_0\pi$
and $\rho\pi$ amplitudes for three values of our parameter $\gamma$.  The value 
$\gamma=-0.08$ is, in some sense, our ``central" value (subject to more refined analyses).
We also plot the phase for a smaller value  $\gamma=-0.95$ and for a larger value $\gamma=-0.065$.  For all three, we observe the spectacular rise  of the phase due to the structure of the P-wave Deck amplitude.  Notice that the larger value $\gamma=-0.065$ seems to be close to a limit beyond which our calculations develop unwanted  effects.  On the other side, for smaller values  $\gamma= < -0.1$, the phase tends to flatten and to be less dramatic in the transition region.  Curiously, for all sensible values of $\gamma$, all curves seem to cross at the same point near $65^\circ$ at a mass of $\sim 1.25$~GeV.\\

Empirically, it may be that the rise of the phase occurs somewhat too soon as compared to the data.  On the other hand, ours is a \emph{first order} calculation of the P-wave $f_0\pi$ amplitude.  The expansion parameter is of the order of $0.1$ so that we can expect deviations of at least $10\%$ in further calculations, which we can perform as needed.\\

Finally, an analysis such as this one can pin down the branching ratio for the $a_1$ decay into $f_0 \pi$. 
A precise value would come out of  a more quantitative fit to data.  It depends strongly on the factor 
$\gamma$, and, for the time being, we prefer to state that the branching ratio
(as compared to the dominant decay mode $a_1\rightarrow \rho\pi$) is of the order of $10^{-3}$.

\section{Concluding remarks}
\label{sec:summary}

Our initial motivation for this study came from the fact that we found it implausible that two $a_1$ resonances could be so close in mass. 
This question led us to a new investigation of coupled channel rescattering effects in the 
Deck diffractive dissociation model.  We were fortunate to rediscover in this 
example that, although a peak is often related to a resonance,  its exact position depends also on the dynamics of the production mechanism by which it was created. The present case is interesting since it is the same diffraction dissociation process that is at work, but with radically different behaviors in different orbital angular momentum states of the dissociated products.  Now that the calculation has been performed, this result seems quite reasonable, if not obvious, but we did not expect such important quantitative effects as a 160 MeV mass difference between two peaks that correspond to the same resonance.\\

One might question whether it would nevertheless be possible to have, within this theoretical framework, two independent $J^P=1^{++}$ resonances so close to each other.  At the onset of this project we did try the hypothesis of two resonances.  But while the two-peak structure is easy to obtain, the phase difference is not.  We had not yet understood fully the mechanism of the P-wave in the Deck model.  So, in order to
make this model cope with two resonances and agree with the COMPASS data one would have to be extremely lucky.  First these two resonances should both be strictly coupled to \emph{different channels}: one to $\rho\pi$ and the other to $f_0\pi$.  In other words, some strict selection rule should be at work, otherwise the P-wave effect, that in the present case has been able to ``double" a resonance peak might also act for each individual peak and  predict not two peaks but three or four!\\ 

As a last remark, we have analyzed the high energy production of the $a_1$ in a strong interaction process.
If, idealistically, one were to perform a low energy $\pi\rho$ and $\pi f_0$ elastic scattering experiment, one would observe 
a single resonance peak as a function of the center-of-mass energy, i.e. that of Fig.~(\ref{D0}), at a position of, e.g., M=1.36 GeV with width 
$\Gamma = 0.31$~GeV.\\

\begin{acknowledgments}
We thank Professor Stephan Paul and other members of the COMPASS collaboration for bringing this interesting topic to our attention. 
One of us (JLB) wishes to thank Prof. Khosrow Chadan, Dr. Philippe Boucaud and Dr. Jean-Pierre Leroy, for their considerable help at the Laboratoire de Physique Th\'eorique d'Orsay with the difficult computations involved in this work.  The work of E.~L.~B. at Argonne is supported
in part by the U.S. DOE under Contract No. DE-AC02-06CH11357.
E.~L.~B. warmly acknowledges  the hospitality of the Aspen Center for Physics where part of this research was done in the summer of 2014.  
\end{acknowledgments}

\appendix

\section{Kinematics and the partial wave decomposition of the Deck amplitude}
\label{sec:PWA}

The process under consideration is $a p \rightarrow a^{*} \pi N$. Four momenta of initial and final state particles  are labeled $p$ and $q$ respectively. Three momenta are denoted by bold characters ${\bf{p}}$.  Norms of the three momenta are denoted by normal characters: $p$ and $q$ when no confusion can arise.\\

We use $A$ for the system $( a^{*} \pi)$ and $M$ for its invariant mass, which we also refer to as $M_{A}$ when it it useful.  The invariants of interest are:
\begin{eqnarray}
 s & =&(p_a +p_p)^2= (q_a+q_\pi+q_N)^2\quad,\label{A1}\\
\vspace{3mm}
 s_{\pi N} & = &(q_\pi + q_N)^2 \equiv M^{2}_{\pi N}\quad,\label{A2}\\ 
\vspace{3mm}
 s_{\pi{a^{*}}}&  =& (q_\pi +q_{a^{*}})^2\equiv M^{2}_A\quad,\label{A3}\\
\vspace{3mm}
 s_{{a^{*}}N}&=&(q_{a^{*}} + q_N)^2\quad, \label{A4}\\
\vspace{3mm}
 t_{a{a^{*}}}&=&(q_{a^{*}}-p_a)^2\quad,\label{A5}\\
\vspace{3mm}
  t_1\equiv t_{pN}&=& (q_N-p_p)^2\quad,\label{A6}\\
\vspace{3mm}
 t_{a\pi}&=& (q_\pi-p_a)^2\quad. \label{A7}
\end{eqnarray}

Note that 
\begin{equation}
s_{\pi{a^{*}}}+t_{a{a^{*}}}+t_{a\pi}=m^2_a+m^2_{a^{*}}+m^2_\pi+t_{pN}\quad,
\label{A8}
\end{equation}

and
\begin{equation}
 s_{\pi {a^{*}}} + s_{\pi N}+s_{{a^{*}}N} = s +m^2_{{a^{*}}} +m^2_\pi+m^2_N\quad.
 \label{A9}
\end{equation}

\subsection{t-channel Variables; Rest Frame of $A$}

\begin{figure}
\begin{center}
\includegraphics[width=.4\textwidth]{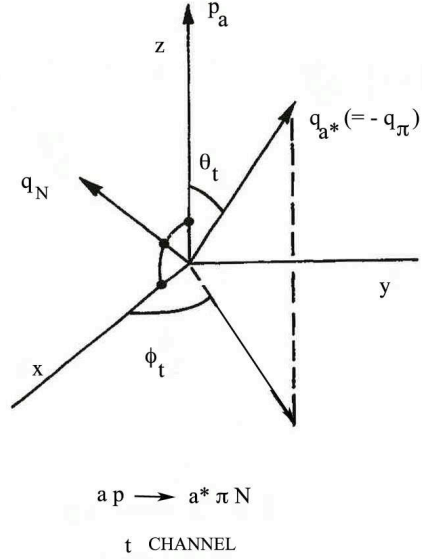}
\end{center}
\caption{t-channel polar angles.}
\label{angles}
\end{figure}

In this calculation, the interesting variables are the set: $s$, $M^2_A$, and
$t_{pN} \equiv t_1$ and the t-channel scattering angles, defined in Fig.~\ref{angles}, in the rest frame of $A$ where ${\bf q}_A= {\bf q}_{a^{*}}+ {\bf q}_\pi= 0$.  In terms of these variables, the energies and magnitudes of the three-vector momenta of $a$, $N$, ${a^{*}}$ and $\pi$ are:
\begin{eqnarray}
 E_a &  = & (M^2_A+ m^2_a-t_1)/ 2M_A \quad,\label{A10}\\
\vspace{3mm}
 p_a & = &\lambda^{1/2}(M^2_A, m^2_a,t_1)/ 2M_A\equiv \vert {\bf p_a}\vert\quad,\label{A11}\\ 
\vspace{3mm}
 E_N  &=& (s - m^2_N-M^2_A)/2M_A\quad,\label{A12}\\
\vspace{3mm}
 q_N & = &\lambda^{1/2}(s, m^2_N, M^2_A)/ 2M_A\equiv \vert {\bf q_N}\vert\quad, \label{A13}\\
\vspace{3mm}
 E_{a^{*}} & = &(s - m^2_N-m^2_\pi)/2M_A\quad,\label{A14}\\
\vspace{3mm}
 \vert {\bf q_\pi}\vert& = &\vert {\bf q_{a^{*}}}\vert=\lambda^{1/2}(M^2_A, m^2_{a^{*}}, m^2_\pi)/ 2M_A\quad,\label{A15}\\
\vspace{3mm}
  E_\pi & = & M_A-E_{a^{*}}\quad\label{A16}, 
\end{eqnarray}

where $$\lambda(A^2, B^2, C^2)= [(A+B)^2-C^2] [(A-B)^2-C^2]\quad.$$

Expressed in terms of polar angles,  Eq. (A.5) becomes 
\begin{equation} 
t_{a{a^{*}}}=m^2_{a^{*}}+m^2_a -2E_{a^{*}}E_a+ 2\vert {\bf q_{a^{*}}}\vert\vert {\bf p}_{a}\vert \cos \theta_t \quad. \label{A18}\\
\end{equation}
The relation of the angle $\chi$ between $\bf{p}_a$ and $\bf{q}_N$, and the polar angles represented in Fig.~(\ref{angles}), is 
\begin{equation}
\cos \psi= \cos \theta_t \cos \chi+ \sin \theta_t \sin \chi \cos \phi_t\quad,
\label{A19}
\end{equation}
where $\psi$ is the angle between $\bf{q}_{a^{*}}$ and $\bf{q}_N$.

\subsection{Amplitudes in the Deck model}

As can be inferred from Eq.~(\ref{DCK}) the quantities of interest in this analysis originate from the ratio
\begin{equation} 
\frac{s_{\pi N}}{t_2-m^2_\pi} \quad, 
\label{ratio}
\end{equation}
where now we specialize to the case of an incident pion ($m_a=m_\pi$)).  Our aim is to obtain an expansion of this amplitude to first order in 
$M^2/s$ and in $t_1/M^2$. The following dimensionless quantity occurs repeatedly:
\begin{equation}
\Theta_1 = \frac{t_1}{(M^2-m^2_\pi)}\quad. 
\label{theta}
\end{equation}
The following expansions are useful:
\begin{eqnarray}
\vspace{3mm}
 E_a &=& \frac{(M^2 + m^2_\pi - t_1)}{2M} \nonumber, \quad\\
 p_a &\approx& \frac{M^2-m^2_\pi}{2M}\left(1 - \frac{(M^2+m^2_\pi)}{(M^2-m^2_\pi)} \Theta_1\right) \nonumber. \quad
 \label{Epa}\\
\end{eqnarray}
\begin{eqnarray}
E_N &= &\frac{s}{2M} (1- \frac{M^2+m^2_N}{s}) \nonumber, \quad\\
q_N &\approx& \frac{s}{2M}(1 - \frac{M^2+m^2_N}{s}) \quad.
\label{EpN}
\end{eqnarray}

Consider the numerator of Eq.~\ref{ratio}. We have:
\begin{equation}
 s_{\pi N}  = (q_\pi + q_N)^2 = m^2_N + m2_\pi+ 2E_\pi E_N - 2q_\pi q_N\cos \psi\quad, 
 \label{spin}
\end{equation}
where $\psi$ is the angle between $\bf{q}_N$ and $\bf{q}_\pi$, as given by Eq.~(\ref{A19}). 

To calculate the angle $\chi$ between $\bf{q}_N$ and $\bf{p}_a$ as defined by Eq.~(\ref{A19}), one can make use of the identity:
\bea
&&(q_{N}-p_a)^2 + (q_{N}+q_{a^{*}}+ q_\pi)^2+ (q_{N}-p_p)^2= \nonumber \\
&&(q_{N}-p_a)^2 + s +t_1=m^2_N+m^2_p+ m^2_\pi+ M^2\quad.
\eea
One obtains, to first order in both $t_1/M^2$ and $M^2/s$, 

\begin{equation}
\cos \chi = -\left(1 + \Theta_1\frac{2M^2 }{(M^2-m^2_\pi)}- \frac{(M^2+m^2_N)}{s}\right),
\label{coski}
\end{equation}
where the quantity $\Theta_1$ has been defined in Eq.~(\ref{theta}).
Since we perform this calculation at the experimental lower value of the momentum transfer, i.e. $t_1 = -0.1$~GeV$^2$, whereas the experiment is performed at $s= 380$~GeV$^2$, the last term on the right is negligible compared to the second one, and this feature remains true in all the calculation. We need to consider only the first order terms in $t_1/M^2$.  It is understandable that $\cos \chi$ should be close to $-1.$ since we consider the forward direction in high energy production.

Using Eq.~(\ref{A19}), we therefore obtain 
\begin{equation}
 s_{\pi N}  = \frac{s}{M}\left( E_\pi - p_\pi \cos \theta (1 + \Theta_1\frac{2M^2 }{(M^2-m^2_\pi)})\right)
 \label{spin2}
\end{equation}   
Notice that we have not taken into account the second term on the right of Eq.(\ref{A19}) since it does not contribute to the partial waves we are interested in (we consider only the $m=0$ amplitudes).\\

The denominator of Eq.~(\ref{ratio}) is simpler.  One obtains from Eq.~(\ref{A18})
\begin{equation}
t_2-m^2_\pi = (p_a-q_{a{^*}})^2-m^2_\pi= m^2_{a{^*}}-2E_aE_{a{^*}}+2p_a p_\pi \cos\theta\quad.
\label{denom}
\end{equation}

The presence of a non vanishing value of $t_1$ can be read off from Eq.~(\ref{Epa}).  One obtains:

\bea
&&t_2-m^2_\pi = - \frac{M^2-m^2_\pi}{M} \times \nonumber \\ 
&&\left( (E_\pi - \Theta_1 E_{a{^*}})- p_\pi \cos \theta (1 -\Theta_1 \frac{(M^2+m^2_\pi)}{(M^2-m^2_\pi)}) \right)\quad.
\label{propag}
\eea

Altogether, the Deck amplitude of interest Eq.~(\ref{ratio}) to first order in $t_1$ is 
\bea
&&T^{Deck} = \frac{s_{\pi N}}{t_2-m^2_\pi} = -\frac{s}{(M^2-m^2_\pi)} \times \nonumber \\
&&\left(\frac{ E_\pi - p_\pi \cos \theta(1 + \Theta_1\frac{2M^2 }{(M^2-m^2_\pi)}) }{ (E_\pi - \Theta_1 E_{a{^*}})- p_\pi \cos \theta (1 -\Theta_1 \frac{(M^2+m^2_\pi)}{(M^2-m^2_\pi)} )}\right)\quad.
\label{ratio2}
\eea

In the limit $\vert t_1 \vert =0$ this expression reduces to $-s/(M^2-m^2_\pi))$ which is a pure S-wave.  All other partial waves 
vanish in the limits $s \to \infty$ and $t_1 \to 0$.  Since Eq.~(\ref{ratio2}) is a rational function of $\cos \theta$, the calculation of 
its partial wave projections is straightforward. 

\subsection{Partial wave projections}
For convenience, we set:
\bea
a  &=&  1+ \Theta_1\frac{2M^2 }{(M^2-m^2_\pi)} \nonumber \\
b  &=&  1 -\Theta_1 \frac{(M^2+m^2_\pi)}{(M^2-m^2_\pi)} \nonumber,\quad \textrm{and}\quad \\
y  &=&   \frac{p_\pi b}{E_\pi - \Theta_1 E_{a{^*}}}.
\label{ab}
\eea
%
%

The S-wave projection of Eq.~(\ref{ratio2}) is obtained by integration over $\cos\theta$: 
\bea 
&& T^{Deck}_S = -\frac{1}{2}\frac{s}{(M^2-m^2_\pi)} \times \nonumber \\
&& \bigg( \frac{2a}{b} -\frac{\Theta_1}{E_\pi - \Theta_1 E_{a{^*}}}\left(E_\pi\frac{(3M^2+m^2_\pi)}{b(M^2-m^2_\pi)}- \frac{a}{b}E_{a{^*}}\right) \times \nonumber \\
&& (\frac{1}{y}\ln\frac{1+y}{1-y} ) 
\bigg) \quad.
\label{swav}
\eea

With the same notation, the P-wave projection of Eq.~(\ref{ratio2}) is obtained after multiplying by $\cos\theta$ and integrating over $\cos\theta$: 
\bea 
&& T^{Deck}_P  =  +\frac{3}{2}\frac{s}{(M^2-m^2_\pi)} \times \nonumber \\
&& \frac{\Theta_1}{(E_\pi - \Theta_1 E_{a{^*}})}\left(E_\pi\frac{(3M^2+m^2_\pi)}{b(M^2-m^2_\pi)}- \frac{a}{b}E_{a{^*}}\right) \times \nonumber \\
&&(\frac{-2}{y} +\frac{1}{y^2}\ln(\frac{1+y}{1-y}))
\quad.
\label{pwav}
\eea
%

Notice that the above expressions are strictly the S- and P-wave projections of Eq.~(\ref{ratio2}).  The terms of order $(\Theta_1)^2$ are incomplete since terms of that order would also occur in Eq.~(\ref{ratio2}).  Therefore, to first order in $\Theta_1$ the S-wave amplitude is:
\bea
&& T^{Deck}_S  = -\frac{s}{(M^2-m^2_\pi)} \times \nonumber \\
&& \left( 1 - \frac{\Theta_1}{2} \left(\frac{(3M^2+m^2_\pi)}{(M^2-m^2_\pi)}- \frac{E_{\rho}}{E_\pi}\right)(\frac{1}{y}\ln(\frac{1+y}{1-y}))\right) \quad; 
\label{swave}
\eea
and $y={p_\pi}/{E_\pi}$.\\

The P-wave amplitude is
\bea
&& T^{Deck}_P  =  +\frac{3}{2}\frac{s}{(M^2-m^2_\pi)}{\Theta_1} \times \nonumber \\
&& \left(\frac{(3M^2+m^2_\pi)}{(M^2-m^2_\pi)}- \frac{E_{f_0}}{E_\pi}\right) (\frac{-2}{y} +\frac{1}{y^2}\ln(\frac{1+y}{1-y})). \quad 
\label{pwave}
\eea


\end{document}